\documentclass[preprint,3p,times]{elsarticle}

\usepackage{lineno,hyperref}
\modulolinenumbers[1]

\usepackage[acronym]{glossaries}

\usepackage{siunitx}
\DeclareSIUnit{\u}{u}
\DeclareSIUnit{\samples}{Sa}
\usepackage {subcaption}
\usepackage[version=4]{mhchem}

\newacronym{MMIC}{MMIC}{monolithic microwave integrated circuit}
\newacronym{HF}{HF}{high-frequency}
\newacronym{RF}{RF}{radio frequency}
\newacronym{DDS}{DDS}{dose delivery system}
\newacronym{SNR}{SNR}{signal-to-noise-ratio}
\newacronym{PCB}{PCB}{printed circuit board}
\newacronym{FWHM}{FWHM}{full width at half maximum}

\journal{NIM A}

\begin{document}

\begin{frontmatter}

\title{Characterizing the Delivered Spill Structure of Medical Proton and Carbon-Ion Beams at MedAustron using a High Frequency Silicon Carbide Readout}

\author[tu_address]{Matthias Knopf \corref{mycorrespondingauthor}}
\cortext[mycorrespondingauthor]{Corresponding author}
\ead{matthias.knopf@tuwien.ac.at}

\author[hephy_address]{Andreas Gsponer}
\author[maus_address,tu_address]{Matthias Kausel}
\author[hephy_address]{Simon Waid}
\author[hephy_address]{Sebastian Onder}
\author[hephy_address]{Stefan Gundacker}
\author[hephy_address]{Daniel Radmanovac}
\author[maus_address]{Giulio Magrin}
\author[hephy_address]{Thomas Bergauer}
\author[tu_address]{Albert Hirtl}

\address[tu_address]{TU Wien, Atominstitut, Stadionallee 2, 1020 Wien}
\address[hephy_address]{Institute of High Energy Physics, Austrian Academy of Sciences, Nikolsdorfer Gasse 18, 1050 Vienna, Austria}
\address[maus_address]{EBG MedAustron GmbH, Marie-Curie Straße 5, 2700 Wiener Neustadt, Austria}

\begin{abstract}
Medical synchrotrons are often used for testing instrumentation in high-energy physics or non-clinical research in medical physics. In many applications of medical synchrotrons, such as microdosimetry and ion imaging, precise knowledge of the spill structure and instantaneous particle rate is crucial. Conventional ionization chambers, while omnipresent in clinical settings, suffer from limitations in charge resolution and integration time, making single-particle detection at high dose rates unfeasible. To address these limitations, we present a beam detection setup based on a silicon carbide (SiC) sensor and a monolithic microwave integrated circuit (MMIC), capable of detecting single particles with a full width at half maximum (FWHM) pulse duration of \qty{500}{\pico\second}. At the MedAustron ion therapy center, we characterized the spill structure of proton and carbon-ion beams delivered to the irradiation room beyond the timescale of the maximum ion revolution frequency in the synchrotron. The resulting data offer valuable insights into the beam intensity at small time scales and demonstrate the capabilities of SiC-based systems for high-flux beam monitoring.
\end{abstract}

\begin{keyword}
Silicon carbide, beam monitor, high dose rate, medical synchrotron, spill structure, microdosimetry
\end{keyword}

\end{frontmatter}


\section{Introduction}

Radiotherapy using light ions has been recognized as a promising option for the treatment of several types of cancer, due to the targeted dose delivery and decreased adverse effects on surrounding tissue ~\cite{Durante_2017_Particle_Therapy, Kim_2020_Carbon_Therapy}. Efforts are being made to further improve the quality of ion beam therapy. This includes, among others, research on higher precision tumor targeting using ion imaging ~\cite{Ulrich_Pur_2020_Ion_Imaging} as well as the adoption of microdosimetry ~\cite{Magrin_2023_State_of_the_Art} into clinical routines for improved dose control. Further, medical accelerators provide a testing platform for several research areas, including detector development for particle physics ~\cite{Waid_2023_Detectors_Particle_Physics} and radiation studies on space electronics ~\cite{Trebersburg_2024_Space}. All these applications rely on the effects of single particles and thus require precise knowledge of the exact spill characteristics at single-particle resolution.\\

Traditionally, in medical accelerator facilities, dosimetry with single-particle precision is not required and challenging to implement at common clinical particle rates up to \SI{e10}{\per\second} ~\cite{ICRU_93_Ion_Therapy}. Since reliability and durability of the equipment are the highest priority, beam monitoring is typically performed using gas-filled ionization chambers to measure the beam intensity \cite{Giordanengo_2013_CNAO_IC}, while segmented ionization chambers are employed to monitor the beam spot position and lateral profile ~\cite{Braccini_2015_Segmented_IC}. These systems are designed to measure the macroscopic dose delivered during treatment in accordance with medical specifications, but lack the precision and speed required for certain experiments. The limited sensitivity of ionization chambers—requiring a minimum charge deposition of several \SI{100}{\femto\coulomb} ~\cite{Giordanengo_2015_CNAO_DDS}—combined with long charge collection times on the order of \SI{100}{\micro\second} ~\cite{Giordanengo_2017_DDS_Concept}, makes them unsuitable for measurements with high temporal resolution, single-particle detection, or low-intensity beam monitoring ~\cite{Ulrich_Pur_2021_Low_Flux}. To meet the demand for higher timing and charge resolution, alternative systems based on solid-state detectors have been proposed ~\cite{Weber_2022_HV_CMOS_Monitor, Data_2024_4D_Tracking}. Solid-state particle sensors offer single particle sensitivity paired with short charge collection times ($<$ \SI{1}{\nano\second}) and can be finely segmented into pixel or strip sensors. Traditionally, silicon is employed as a sensor material. However, concerns about radiation damage and temperature stability warrant the search for novel detector materials. Silicon carbide (SiC) is a wide-bandgap semiconductor, which has recently gained renewed interest as a detector material for applications in future particle physics experiments, fusion research, and space applications ~\cite{DeNapoli_2022_SiC}. With a bandgap of \SI{3.26}{\electronvolt}, SiC devices exhibit very low dark currents compared to silicon detectors. This remains true even after exposure to high radiation fluences, hinting at a superior radiation hardness of SiC detectors compared to silicon devices ~\cite{Gsponer_2023_Neutron}. Additionally, the high charge carrier velocities in combination with a high breakdown field render SiC suitable for fast readout and timing applications ~\cite{Yang_2022_SiC_Time_Res}.  Diamond detectors have also been proposed for beam monitoring \cite{Frais_2007_CVD_Beam_Monitor}, as they offer a large bandgap, good temperature stability, and high radiation hardness. However, their high cost, limited sample sizes, and other manufacturing constraints have encouraged the use of SiC as an alternative detector material ~\cite{Medina_2023_SiC_vs_Diamond}.  This unique set of properties makes SiC an ideal detector material for beam monitoring purposes at high dose rates. SiC-based systems have demonstrated the capacity to monitor the beam current at clinical intensities with minimum dark current ~\cite{Waid_2024_HDM1, HDM2}, as well as for ultra-high dose rates~\cite{waid_pulsed_2024, fleta_2024_SiC_FLASH}.\\

Combining low-capacitance SiC diodes with low-noise \gls{HF} amplifiers enables excellent timing resolution. The low capacitance is provided by a small active area, which also reduces the average number of particles hitting the sensor, reducing pileup to acceptable levels at medical beam intensities. In this work, we expand upon SiC-based beam monitoring efforts by research teams at MedAustron ~\cite{waid_pulsed_2024, Waid_2024_HDM1, HDM2} by analyzing spills on a particle-by-particle basis.  Using a SiC-based \gls{HF} readout system, we present measurements of proton and carbon-ion beams at clinically relevant energies and dose rates on a small surface area detector with high temporal resolution. This setup allows detailed spill structure analysis in the treatment room, offering valuable insights for researchers using the extracted ion beams. This is particularly relevant in applications such as microdosimetry, where the microscopic spill structure affects the quality of the data due to pulse pileup. The results of the measurements further emphasize the potential of SiC-based detector systems for next-generation beam monitoring and provide input for the potential future development of a spatially resolved readout covering the whole beam spot with a segmented detector. 

\section{Materials \& Methods}

\subsection{The MedAustron facility}

MedAustron is an ion therapy center, located in Wiener Neustadt (Austria), featuring four clinical beamlines as well as a dedicated research beamline. The facility is centered around a medical synchrotron, based on the proton-ion medical machine study (PIMMS) ~\cite{PIMMS}, with \SI{77.65}{\meter} circumference, which allows for the acceleration of protons from \SIrange{62.4}{800}{\MeV},\ce{^12C^6+} ions from \SIrange{120}{402.8}{\MeV\per\u} ~\cite{Pivi_2019_Status_Carbon_Commissioning} and \ce{^4He^{2+}} ions from \SIrange{39.8}{402.8}{\MeV\per\u} ~\cite{Gambino_2024_Helium}. The MedAustron design is optimized to fulfill medical requirements. In clinical operation, the accelerated ions are extracted via betatron core-driven third-order resonant extraction ~\cite{Pullia_2016_Betatron_Extraction}. The particles are extracted in spills lasting several seconds, which corresponds to a large number of revolutions in the synchrotron. Particle rates obtained in the irradiation rooms are typically in the range from \SIrange{e9}{e10}{\per\second} for protons and \SIrange{e8}{e9}{\per\second} for carbon-ions ~\cite{Pullia_2000_Slow_Extraction, Benedikt_2010_Overview_MAUS}. Due to the influence of power converters on the magnetic fields in the synchrotron, the extracted spill exhibits significant fluctuations in intensity (ripples) in the \SI{}{\kilo\hertz} region ~\cite{PIMMS}. To mitigate these ripples and ensure a constant particle flux for treatment, MedAustron employs empty bucket channeling ~\cite{Kühteubl_Diss, Pullia_2016_Betatron_Extraction}. While this technique achieves a reduction of the intensity ripples, it modulates the spill structure at the timescale of the ion revolution frequency in the synchrotron ($\sim$ \SIrange[]{1}{3}{\mega\hertz}). In clinical treatment, this modulation can be neglected \cite{Crescenti_1998_EBC}. However, it proves to be relevant for applications relying on single-particle precision.

\subsection{SiC-based \gls{HF} readout}

The sensor employed in the detector system was a 4H-SiC p-i-n diode designed by the authors and manufactured by IMB-CNM-CSIC ~\cite{CNM}. In order to achieve readout at high bandwidth (\(> \SI{1}{\giga\hertz}\)), the detector capacitance had to be kept at a minimum. Thus, a small-area circular sensor-design with \SI{140}{\micro\meter} diameter and \SI{50}{\micro\meter} thickness was chosen for the experiment. At full depletion (\(>\SI{150}{\volt}\)), the measured capacitance was \SI{70}{\femto\farad}. The sensor consists of an n-doped active region epitaxially grown on a \SI{350}{\micro\meter} substrate ~\cite{Gsponer_2023_Neutron}. The small surface area of \SI{0.015}{\milli\meter\squared} reduces the average particle rate of the ion beams sufficiently to allow for a single-particle detection at medical intensities.  While pileup can never be fully excluded, the likelihood of two or more particles arriving within the specified timing threshold (\SI{1}{\nano\second}) on the small detector is very low.  During the measurements, the detector was operated at \SI{1}{\kilo\volt} reverse bias using a Keithley 2470 SMU.\\

\begin{figure*}[h!]
    \centering

    \begin{subfigure}[b]{0.4\textwidth}
        \centering
        \includegraphics[width=0.99\textwidth]{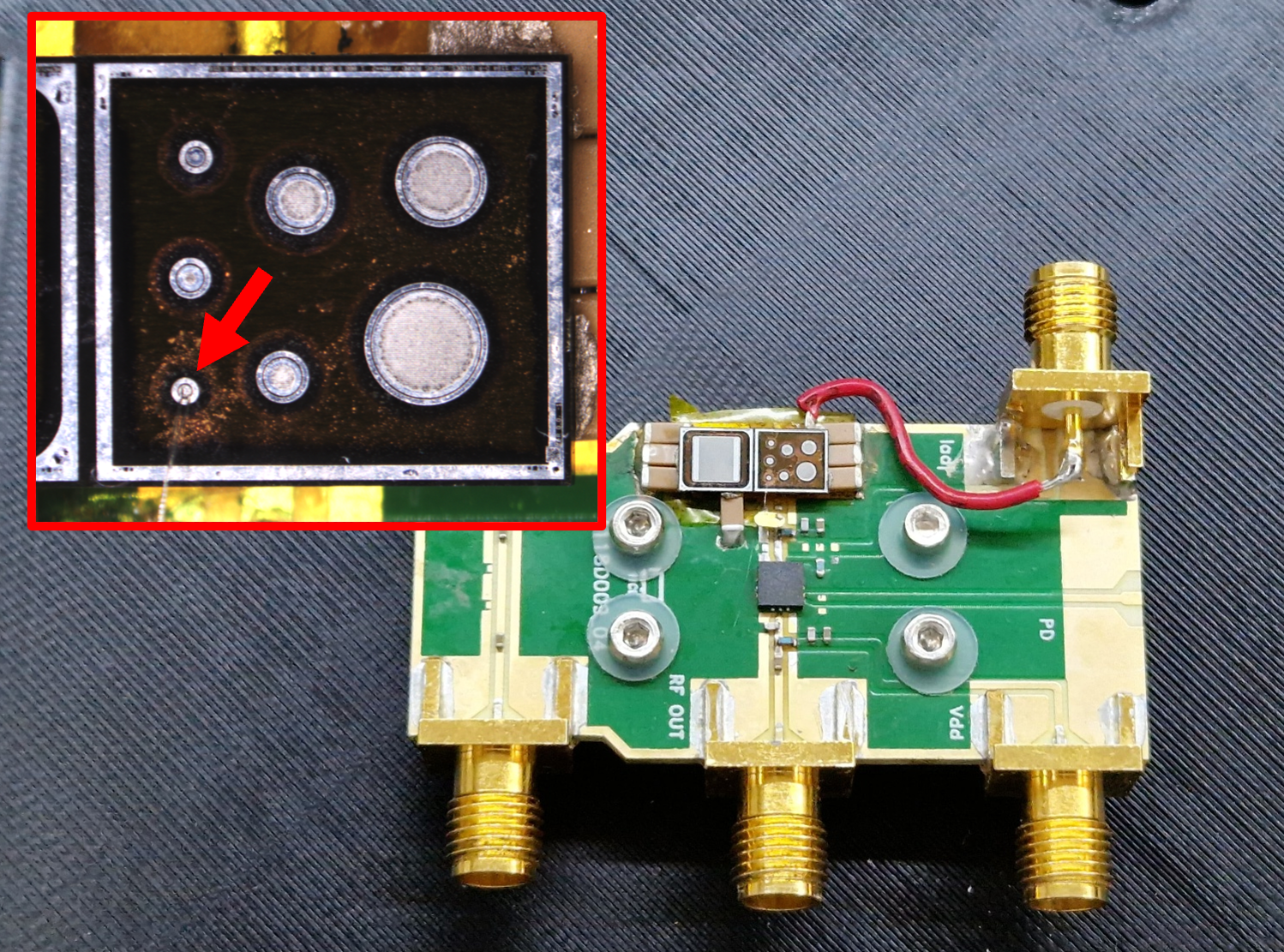}
        \caption{SiC detector and \gls{HF} amplifier mounted on the \gls{PCB}}
        \label{fig:det_amp}
    \end{subfigure}
    \hfill
    \begin{subfigure}[b]{0.55\textwidth}
        \centering
        \includegraphics[width=0.99\textwidth]{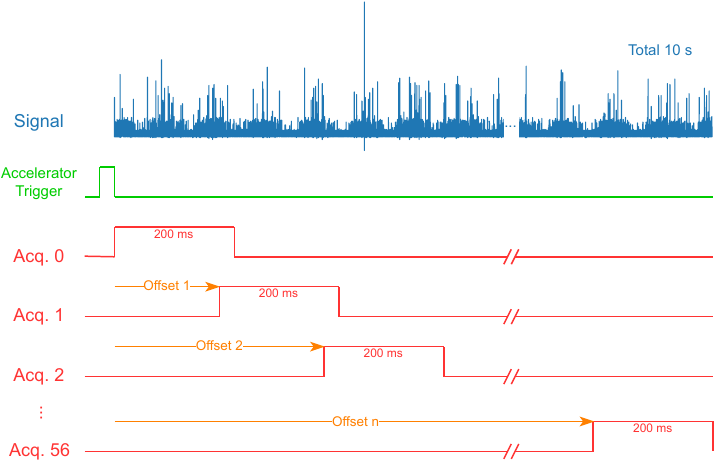}
        \caption{Oscilloscope acquisition scheme  (not to scale) }
        \label{fig:osci_scheme}
    \end{subfigure}

    \caption{a) The readout setup used for the measurements at MedAustron. The detector used is highlighted by the red arrow in the zoomed inset. The commercial \gls{PCB} was modified to optimize readout performance in the multi-\SI{}{\giga\hertz} frequency range and the high bias voltage for the SiC detector. b) Schematic depiction of the acquisition using a digital oscilloscope. A \SI{10}{\second} spill is captured in pieces by shifting the horizontal offset after the extraction start trigger from the accelerator.}
    \label{fig:daq_setup}
\end{figure*}

The sensor readout for the detector system was based on a Mini-Circuits PMA3-14LN+ \gls{MMIC} low-noise amplifier. This amplifier has bandwidth of \SI{10}{\giga\hertz} bandwidth, a gain of \SI{22.6}{\decibel}, and a typical noise figure of \SI{1.1}{\decibel}. A Mini-Circuits PMA3-14LN+ evaluation board was modified to accommodate a first-stage PMA3-14LN+ amplifier, a SiC sensor as well as its high-voltage bias, while minimizing the length of the bond-wires, in order to reduce parasitic inductance and to ensure \SI{50}{\ohm} impedance matching at the amplifier input. Details on the readout can be found in ~\cite{HF_Readout}. Additionally, an SMA-connected Mini-Circuits ZX60-14LN-S+ low-noise amplifier with a gain of about \SI{22}{\decibel} was used as a secondary stage before digitization. A photograph of the SiC radiation sensor and the first amplifier stage is shown in figure \ref{fig:det_amp}.\\

\begin{figure*}[h!]
    \centering

        \includegraphics[width=0.85\textwidth]{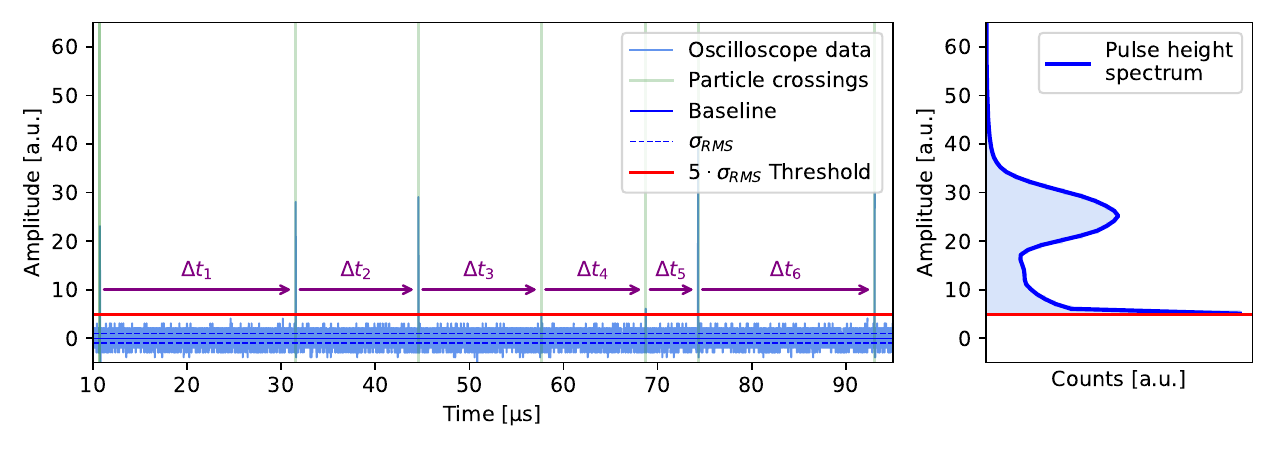}
        \caption{Single particle crossings and their respective inter-arrival times $\Delta t_i$ are registered in the digitized data as threshold crossing above $5\cdot\sigma_\text{RMS}$ noise. The construction of a pulse height spectrum from the peak maxima is indicated.}
        \label{fig:pileup_prob_clinical}

    \label{fig:daq}
\end{figure*}

Digitization was carried out using a Rhode \& Schwarz RTO6 oscilloscope at a sampling rate of \SI{3.33}{\giga\samples\per\second}. Since the internal memory of the oscilloscope is limited to 2 GPts, the \SI{10}{\second} spills were recorded in segments of \SI{200}{\milli\second}. Data acquisition was triggered using the accelerator timing system. The horizontal offset of the oscilloscope was incrementally increased between the recorded segments, with \SI{20}{\milli\second} overlap. This approach allows for time-resolved analysis of the entire detector signal over the full duration of the spill. The process is indicated in figure \ref{fig:osci_scheme}.  Particle crossings were identified in the signal as rising-edge threshold crossings above the RMS noise level, $5\cdot \sigma_\text{RMS}$. The average signal waveform for a particle crossing exhibits a risetime of $\sim$\SI{100}{\pico\second} and a \gls{FWHM} of $\sim$\SI{500}{\pico\second}. Since the signal rise time is shorter than the sampling interval of \SI{300}{\pico\second}, each timestamp directly corresponds to the acquired peak maximum. This results in a timing uncertainty of $\sigma_t=300 \text{ ps} / \sqrt{12}\approx$ \SI{87}{\pico\second}, which is negligible compared to the beam modulation timescale of several hundred \si{\nano\second}. The approach enables a measurement of the inter-arrival time between consecutive particles, as shown in figure \ref{fig:daq}.   To ensure sufficient temporal separation for reliable event discrimination, the minimum accepted interval between consecutive hits was set to \SI{1}{\nano\second}.\\

\subsection{Measurements at MedAustron}

The SiC-based \gls{HF} readout was tested at MedAustron, characterizing proton and carbon-ion spills at medical intensities. Data were collected for the highest and lowest available clinical energies (\SIrange{62.4}{252.7}{\MeV} for protons and \SIrange{120}{402.8}{\MeV\per\u} for carbon-ions) with a spill length of \SI{10}{\second}. A second set of proton and carbon data with reduced particle flux but otherwise unaltered accelerator settings was acquired. To this end, the intensity-degrader in the injector was employed ~\cite{Adler_2019_MAUS_Degrader}. A degradation down to nominally \SI{10}{\percent} of the clinical rate was chosen. During the irradiation, the detector was placed in the isocentre of the MedAustron irradiation room one (IR1). As a reference measurement, the intensity and beam spot position, along with its horizontal and vertical spot size, were monitored on a spill-by-spill basis throughout the entire campaign using the beam delivery system installed  in IR1 . The average measured particle rate was on the order of \SI{e8}{\per\second} for the carbon-ion spills and \SI{e9}{\per\second} for the proton spills.\\

\section{Results}

\subsection{Inter-arrival times}

\begin{figure*}[htp]
    \centering
    \begin{subfigure}[b]{0.48\textwidth}
        \centering
        \includegraphics[width=.99\textwidth]{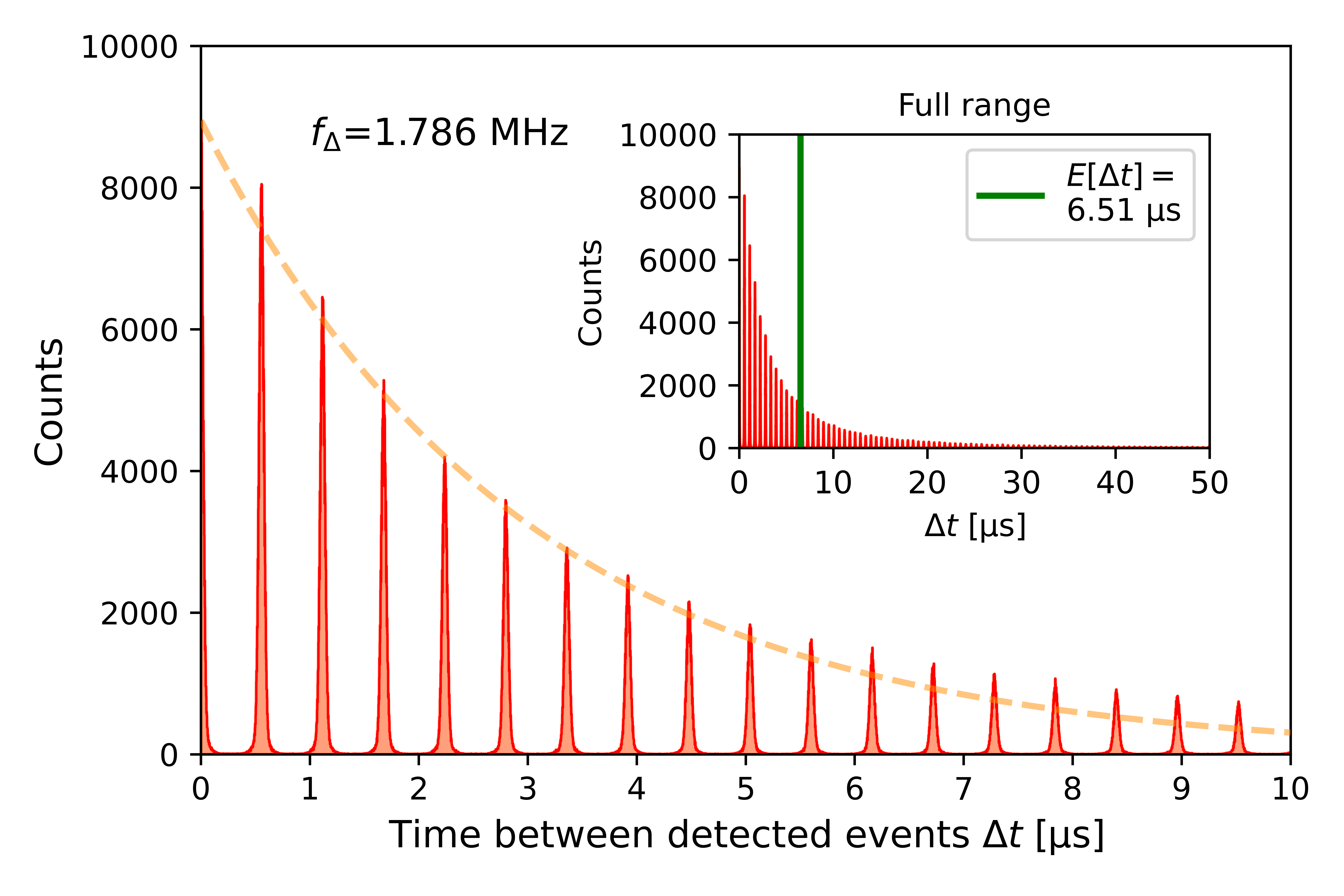}
        \caption{\ce{^12C^6+} ions at \SI{120}{\mega\electronvolt\per\u}}
        \label{fig:hist_120MeV_clinical}
    \end{subfigure}
    \hfill
    \begin{subfigure}[b]{0.48\textwidth}
        \centering
        \includegraphics[width=.99\textwidth]{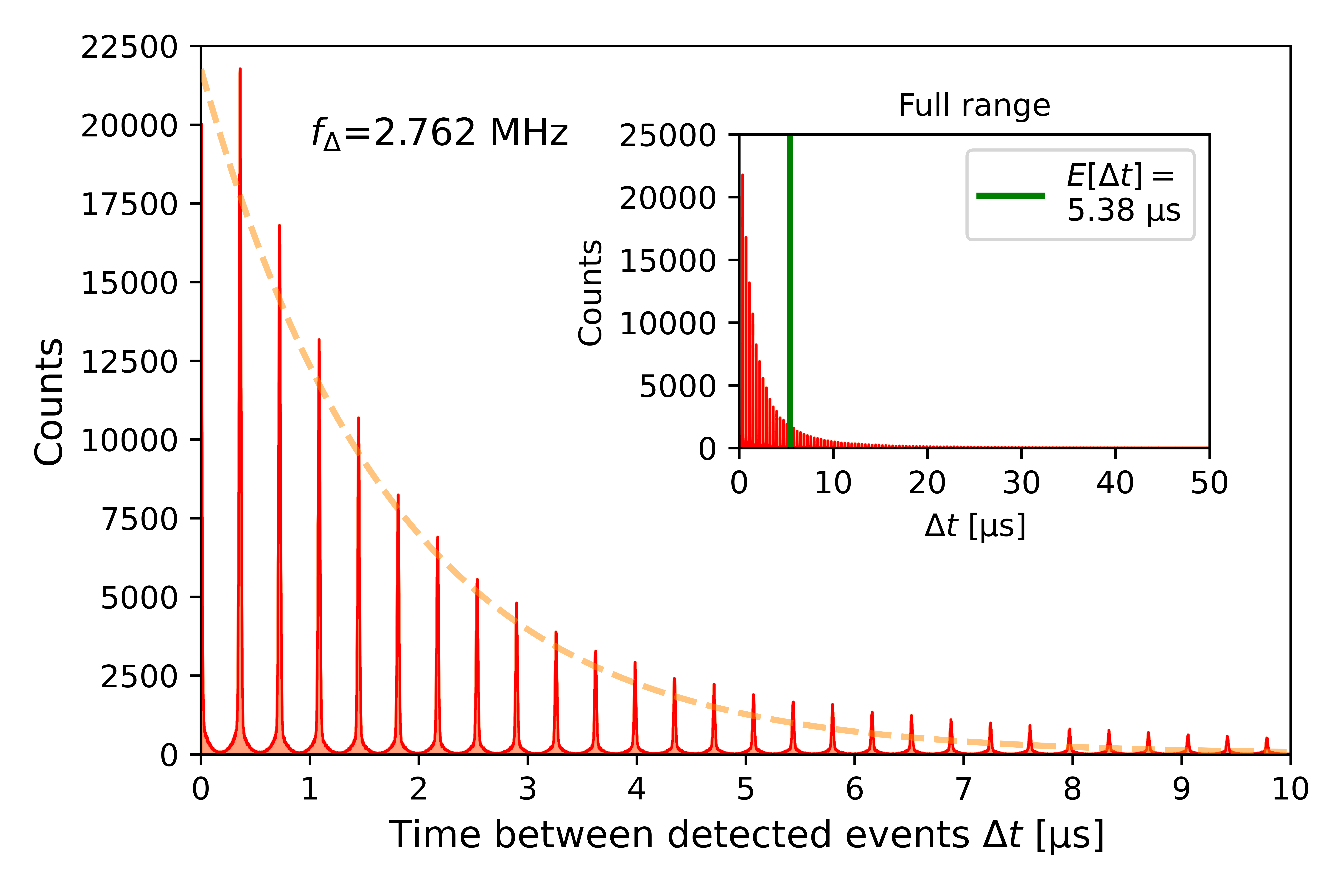}
        \caption{\ce{^12C^6+} ions at \SI{402.8}{\mega\electronvolt\per\u}}
        \label{fig:hist_402MeV_clinical}
    \end{subfigure}
    
    \vspace{0.5cm}
    
    \begin{subfigure}[b]{0.48\textwidth}
        \centering
        \includegraphics[width=.99\textwidth]{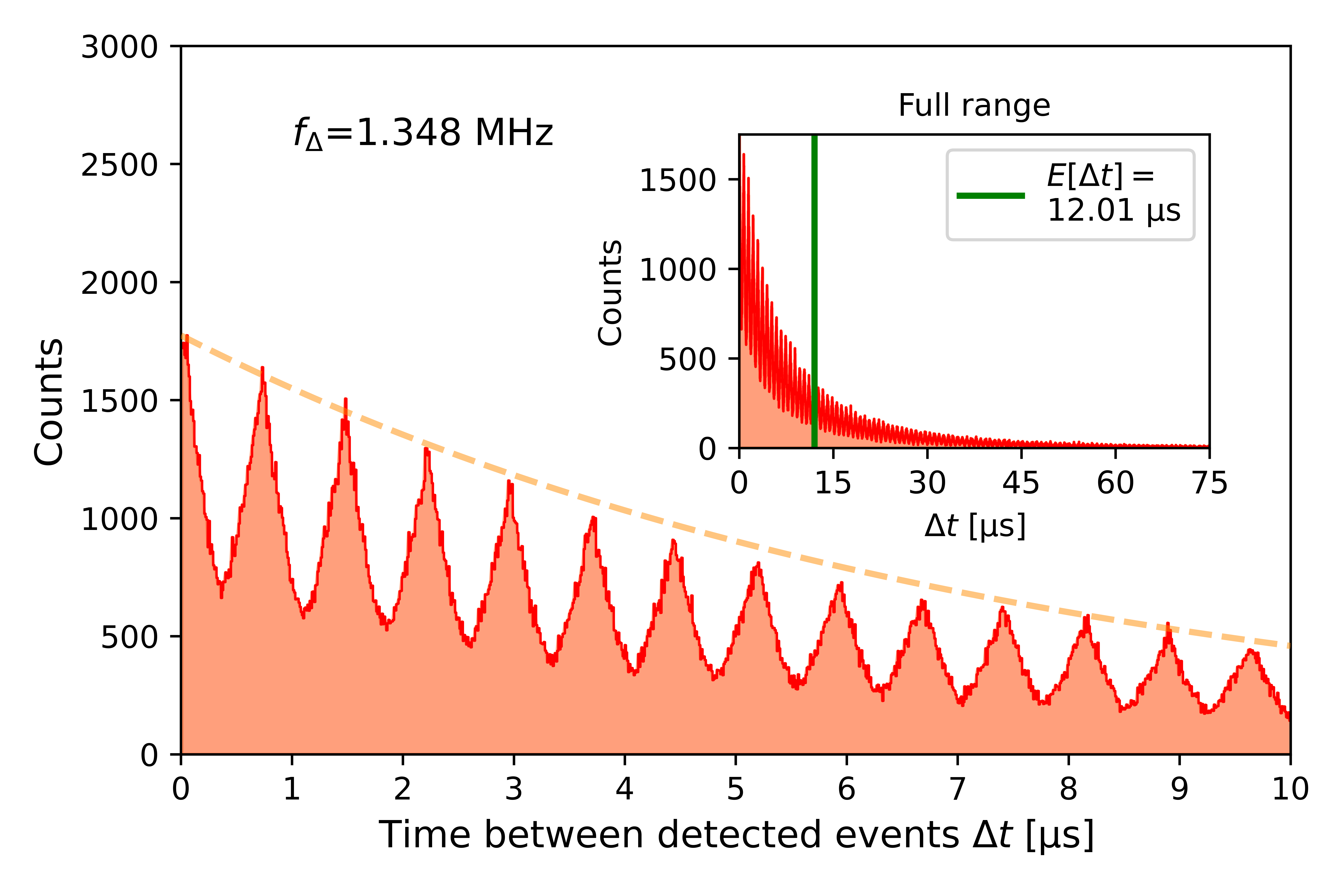}
        \caption{Protons at \SI{62.4}{\mega\electronvolt}}
        \label{fig:hist_p_62MeV_clinical}
    \end{subfigure}
    \hfill
    \begin{subfigure}[b]{0.48\textwidth}
        \centering
        \includegraphics[width=.99\textwidth]{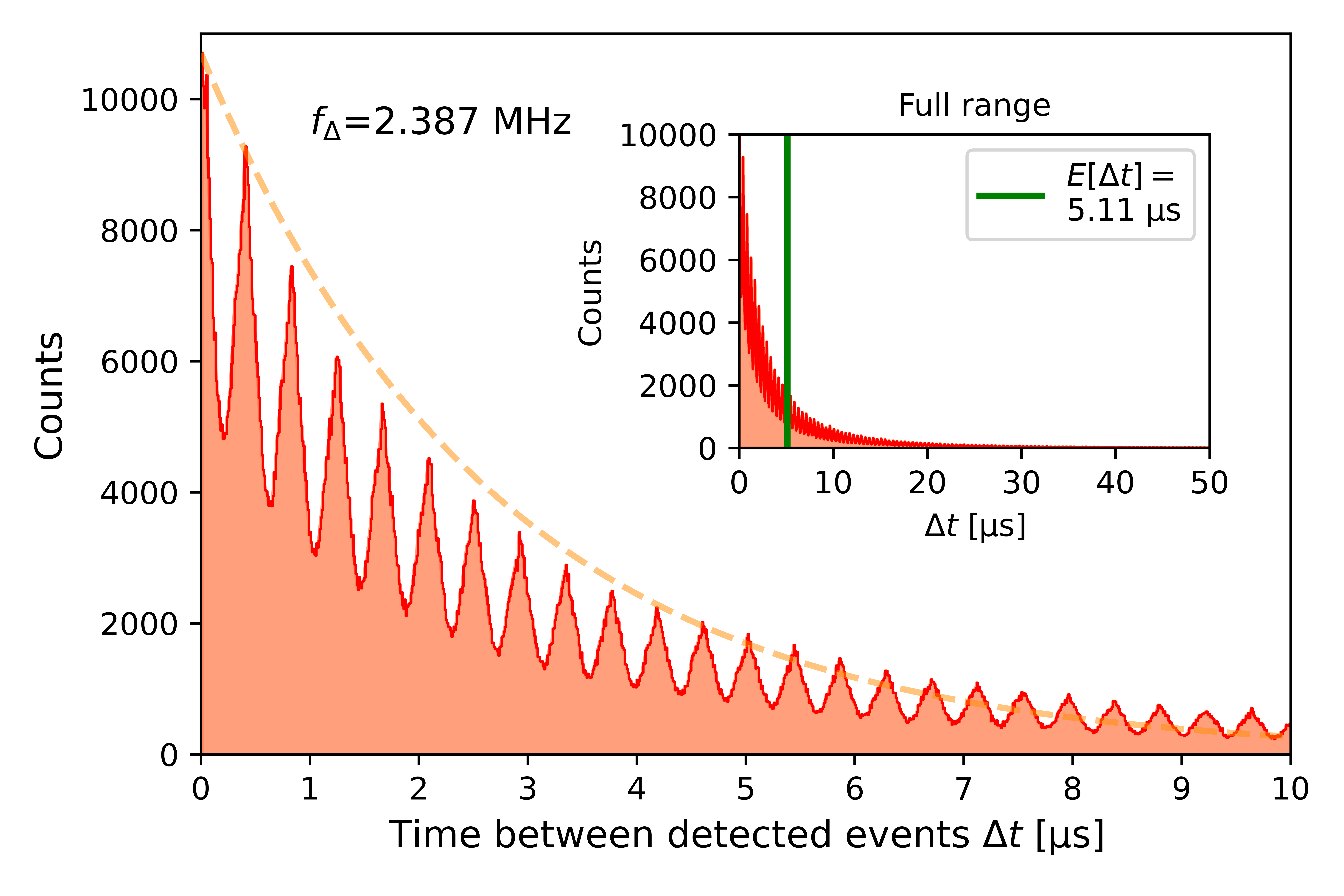}
        \caption{Protons at \SI{252.7}{\mega\electronvolt}}
        \label{fig:hist_p_252MeV_clinical}
    \end{subfigure}

    \caption{Time interval distributions between consecutive detections for proton and carbon-ion beams at clinical rates. The measurements exhibit the characteristic comb-like structure caused by empty bucket channeling. The distributions can be approximated by exponential functions, as shown by the curves determined from least-squares fitting in the plots. The inset in the top right displays the distribution over a wider timescale, with the expectation value $E[\Delta t]$ of the distributions marked in green. The \gls{RF} frequencies determined from the mean peak positions $f_\Delta$ are indicated in the plots.}
    \label{fig:inter_arrival_histogram}
\end{figure*}

Figure \ref{fig:inter_arrival_histogram} shows the distribution of inter-arrival times $\Delta t$ between two consecutive particle detections for proton and carbon-ion beams at clinical intensities. All measurements at full clinical rate and at \SI{10}{\percent} nominal rate show a monotonically decreasing distribution with superimposed, evenly spaced peaks. The lower stopping power of high-energy protons compared to carbon-ions results in a  reduced \gls{SNR}. Nonetheless, pulse height spectra created from the individual particle crossings exhibit the characteristic Landau/Gaussian shape above the $5\cdot \sigma_\text{RMS}$ noise threshold (see figure \ref{fig:daq}) for all carbon measurements and the \SI{62.4}{\mega\electronvolt\per\u} proton measurements, indicating sufficient amplitude resolution. For the \SI{252.7}{\mega\electronvolt\per\u} proton measurements only the Landau tail is observed in the pulse height spectra , since only the highest energy deposits are able to cross the $5\cdot \sigma_\text{RMS}$ threshold.  The spacing of the peaks in the inter-arrival time distribution corresponds to the synchrotron \gls{RF} frequency, given by the ion revolution frequency $f_{\text{RF}} = \beta c / L$, where $\beta$ is the particle velocity and $L$ is the accelerator ring circumference. This characteristic structure results from empty bucket channeling ~\cite{Crescenti_1998_EBC}. In the carbon data, the shape of the peaks is well described by Gaussian functions, allowing precise extraction of their positions from least squares fits. For protons, peak positions are determined by calculating the mean bin position per peak. The frequencies $f_\Delta = 1 / \bar{T_\Delta}$ derived from the mean peak spacing $T_\Delta$ for both carbon-ions and protons agree with the calculated ion revolution frequencies with an average error of \SI{0.25}{\percent}, excluding the lower-flux proton data due to insufficient \gls{SNR}. The data is provided in table \ref{tab:table}. The slope of the declining peak heights can be approximated with exponential functions $f(\Delta t) = \exp{(-\alpha \Delta t)}$. While the initial decay closely follows the exponential model, significant deviations occur in the tail. Least-squares fitting yielded a mean standard error of \SI{1.34}{\percent} for all fits, which was derived from the covariance matrix of the fit. The fits are shown in figure~\ref{fig:inter_arrival_histogram}, and the results are summarized in table~\ref{tab:table}. The expectation value for the inter-arrival distribution $f(\Delta t_i)$ with bins $\Delta t_i$ can be calculated as

\begin{equation}
    E[\Delta t] = \frac{\sum_i \Delta t_i f(\Delta t_i)}{\sum_i f(\Delta t_i)}.
\end{equation}

The expected mean time interval between subsequent particles $E[\Delta t]$ is thus in the \SI{}{\micro\second} range. However, particle bunching was observed for time intervals down to $\Delta t =$ \SI{1}{\nano\second}, which was set as the minimum resolution in our analysis. It is assumed that the exponential behavior extends beyond this limit.  The limited \gls{SNR} in the high-energy proton measurements introduces a selection bias that affects the absolute particle rate recorded. However, this does not lead to systematic errors in the characterization of the beam microstructure. The same effect can be observed in the measurements at \SI{10}{\percent} nominal rate which show the same microscopic modulation as the high-rate measurements, despite differences in the macroscopic decline rate and overall event count. Since the deposited energy of particles with the same kinetic energy varies randomly, only registering higher-amplitude events does not preferentially sample any particular subset of particles within the beam. 

\subsection{Spill structure}
\begin{figure*}[h]
    \centering
    
    \begin{subfigure}[b]{0.48\textwidth}
        \centering
        \includegraphics[width=.99\textwidth]{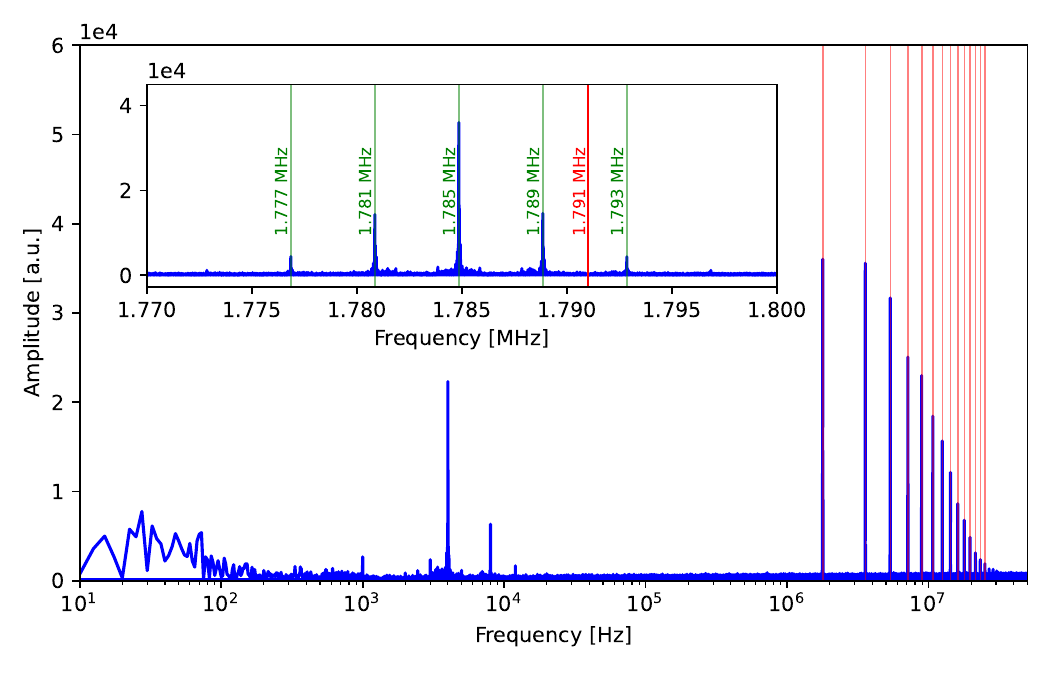}
        \caption{\ce{^12C^6+} ions at \SI{120}{\mega\electronvolt\per\u}}
        \label{fig:FFT_120MeV}
    \end{subfigure}
    \hfill
    \begin{subfigure}[b]{0.48\textwidth}
        \centering
        \includegraphics[width=.99\textwidth]{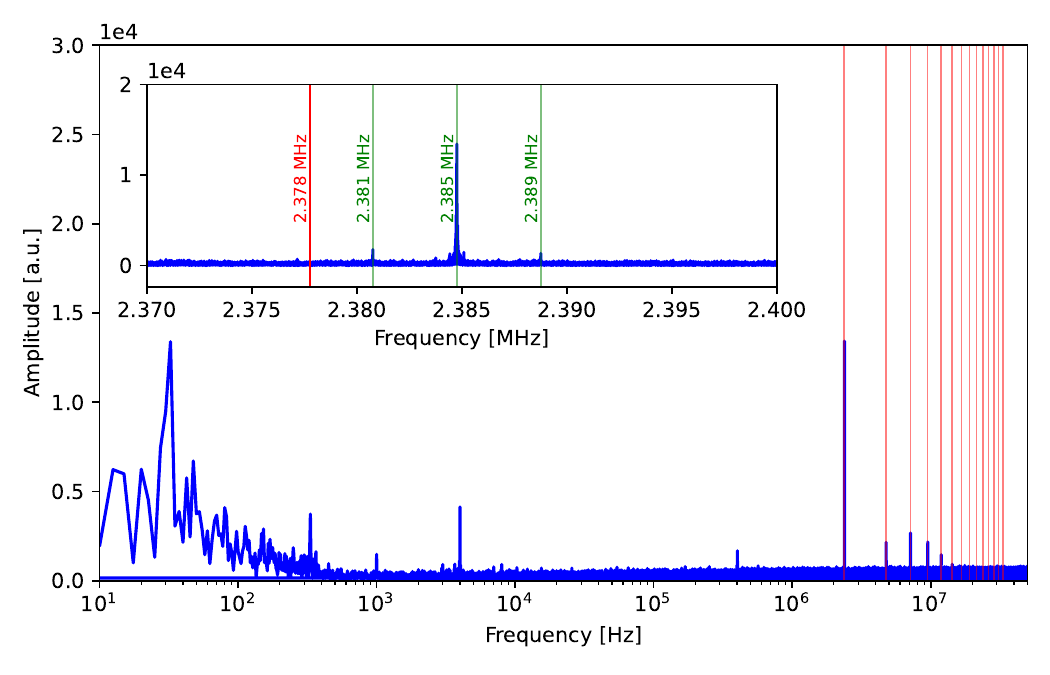}
        \caption{Protons at \SI{252.7}{\mega\electronvolt}}
        \label{fig:FFT_252MeV}
    \end{subfigure}

    \caption{Frequency spectra of the beam intensities for a) carbon-ions and b) protons. Ripples from power supplies appear in the \SI{}{\kilo\hertz} range, while modulation at the synchrotron \gls{RF} frequency (shown in red) is visible in the \SI{}{\mega\hertz} range. The inset shows a zoomed section around the first \SI{}{\mega\hertz} peak, with the revolution frequency $f_\text{RF}$ indicated in red.}
    \label{fig:FFT_Intensity}
\end{figure*}

The beam intensity was extracted from the data by integrating the acquired counts in \SI{20}{\micro\second} bins. The frequency spectra of the beam intensity are shown in figure \ref{fig:FFT_Intensity}. The intensity modulation introduced by the empty bucket channeling are clearly visible in the \SI{}{\mega\hertz} frequency band for both proton and carbon-ion beams. These peaks are offset from the ion revolution frequency $f_\text{RF}$ by several \SI{}{\kilo\hertz} (at the synchrotron \gls{RF} frequency), with each harmonic exhibiting a substructure of secondary peaks spaced by \SI{4}{\kilo\hertz}. Additionally, the characteristic ripples from the power converters at MedAustron are visible in the \SI{}{\kilo\hertz} frequencies, with a main peak emerging at \SI{4}{\kilo\hertz} from the magnet power converters ~\cite{DeFranco_2021_MAUS_Optimization}. These findings are consistent with previous analyses of the MedAustron intensity profile ~\cite{Waid_2024_HDM1, Kühteubl_Diss}. Using the \gls{HF} readout, it is possible to monitor the intensity profile of the spills over the whole relevant frequency range while simultaneously allowing for a micro-structure analysis.\\

\begin{table*}[ht]
    \centering
    \caption{Values extracted from the inter-arrival time distributions. Shown are the calculated synchrotron revolution frequency $f_\text{RF}$, the frequency $f_{\Delta}$ determined from the distances between peaks in the inter-arrival time distribution $T_\Delta$, the expectation value $E[\Delta t]$ of the distribution, and the exponential fit parameter $\alpha$ for measurements at both full clinical rate and for the measurements at \SI{10}{\percent} nominal rate. The uncertainty in $f_\Delta$ reflects the standard deviation of the individual peak intervals determined from the distributions. \label{tab:table}}
    
    \smallskip
    \small
    \begin{tabular}{c|c|c|c|c|c|c}
        \textbf{Particle} & \textbf{Energy [\SI{}{\mega\electronvolt\per\u}]} & \textbf{Nominal rate [\%]} & \textbf{$f_\text{RF}$ [\SI{}{\mega\hertz}]} &  \textbf{$f_{\Delta}$ [\SI{}{\mega\hertz}]} & \textbf{$E[\Delta t]$ [\SI{}{\micro\second}]} & \textbf{Exp. fit $\alpha$ [\SI{}{\per\second}]}\\
        \hline
        protons & 62.4 & 100 & 1.342  & \num{1.348(0.085)} & 12.01 & \num{1.35e5}\\
        protons & 252.7 & 100 & 2.378 & \num{2.387(0.403)} & 5.11 & \num{3.69e5}\\
        protons & 62.4 & 10 & 1.342 & \num{1.292(0.297)} & 57.72 & \num{3.42e4}\\
        protons & 252.7 & 10 & 2.378 & \num{2.212(0.881)} & 48.68 & \num{5.88e4}\\
        
        \hline

        \ce{^12C^6+} ions & 120 & 100 & 1.791 & \num{1.786(0.007)} & 6.51 & \num{3.38e5}\\
        \ce{^12C^6+} ions & 402.8 & 100 & 2.764 & \num{2.762(0.031)} & 5.38 & \num{5.68e5}\\
        \ce{^12C^6+} ions & 120 & 10 & 1.791 & \num{1.786(0.012)} & 53.91 & \num{4.09e4}\\
        \ce{^12C^6+} ions & 402.8 & 10 & 2.764 & \num{2.762(0.039)} & 40.67 & \num{7.05e4}\\

        \hline
    \end{tabular}
\end{table*}

\section{Discussion}

\begin{figure*}[h!]
    \centering
    \begin{subfigure}[b]{0.48\textwidth}
        \centering
        \includegraphics[width=0.99\textwidth]{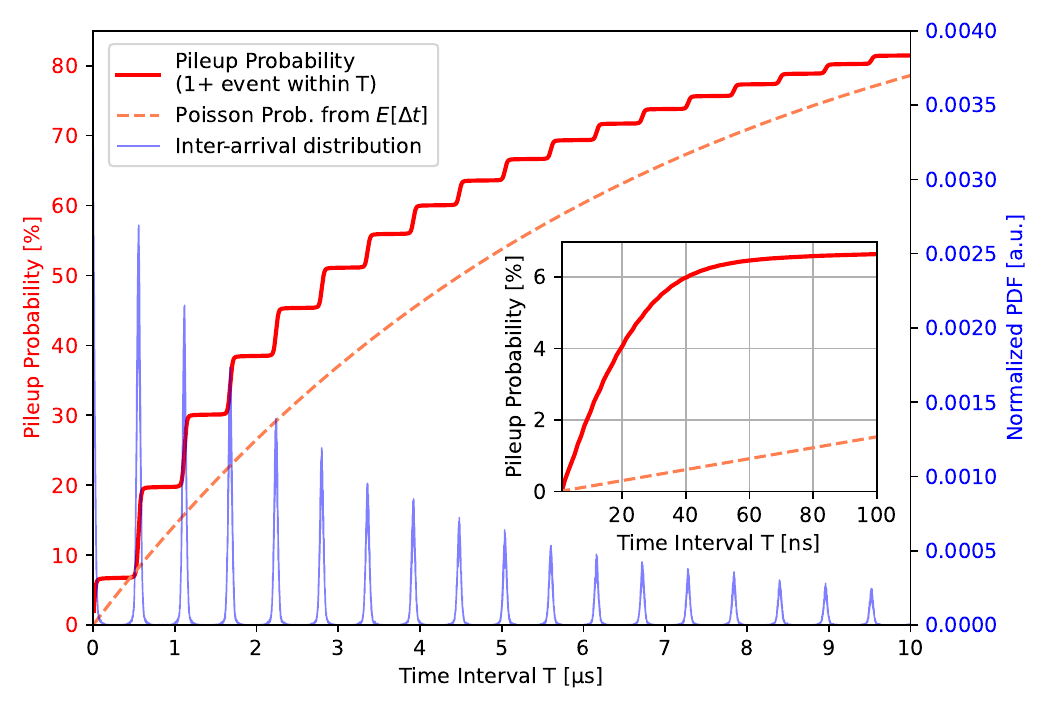}
        \caption{\SI{120}{\mega\electronvolt\per\u} \ce{^12C^6+} ions at clinical intensity}
        \label{fig:pileup_prob_clinical}
    \end{subfigure}
    \hfill
    \begin{subfigure}[b]{0.48\textwidth}
        \centering
        \includegraphics[width=0.99\textwidth]{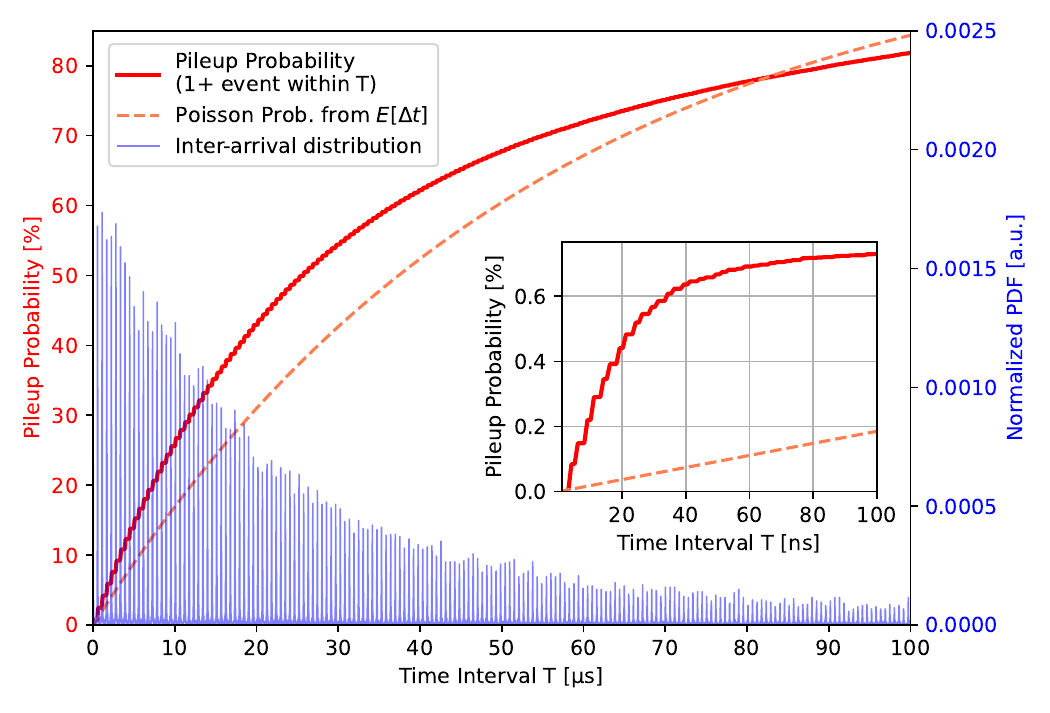}
        \caption{\SI{120}{\mega\electronvolt\per\u} \ce{^12C^6+} ions at clinical intensity}
        \label{fig:pileup_prob_degrader}
    \end{subfigure}

    \caption{The pileup probabilities (red) calculated from the inter-arrival distributions (blue) for \SI{120}{\mega\electronvolt\per\u} \ce{^12C^6+} ions at a) full clinical intensity (\SI{4.01(0.87)e8}{\per\second}) and b) in the accelerator setting at \SI{10}{\percent} nominal intensity (\SI{4.84(0.24)e7}{\per\second}). Pileup is defined as the probability of the next particle detection within a time interval $T$ after the first detection. The respective Poisson pileup probability $1-\exp({E[\Delta t] \cdot T})$, calculated from the expectation value of the inter-arrival distribution is indicated with the dashed line. The inset in the plots shows the pileup probability from \SIrange{1}{100}{\nano\second}.}
    \label{fig:Pileup_prob}
\end{figure*}

A SiC-based \gls{HF} readout system was used to investigate the spill structure of proton and carbon-ion beams at high dose rates on a single-particle level. This work supplements previous efforts in monitoring the beam intensity at MedAustron using SiC detectors ~\cite{Waid_2024_HDM1, HDM2} with an analysis of the micro- and nano-structure of the spills. The presented data may offer important guidance for experimental design at medical synchrotrons, providing insights into beam delivery at the iso-centre. In energy-resolved measurements performed at single-particle resolution, estimating pulse pileup is essential and helps in the appropriate tuning of experimental parameters and acceptable electronic processing times. Neglecting second-order effects, pileup probability can be calculated as the cumulative distribution function of the inter-arrival distributions via piecewise integration. The definition of pileup is thus, the arrival of a second hit within a given time interval $T$ after the first detection. The distribution function was integrated from the detection limit of \SI{1}{\nano\second} up to \SI{1}{\milli\second}, beyond which no additional counts were observed in the data. Figure \ref{fig:Pileup_prob} shows the results for \ce{^12C^6+} ions at \SI{120}{\mega\electronvolt\per\u}. At clinical intensity, an average particle rate of \SI{4.01(0.87)e8}{\per\second} on a Gaussian beam spot with \SI{6.05(0.11)}{\milli\meter} $\times$ \SI{6.18(0.18)}{\milli\meter} \gls{FWHM} was determined from the \gls{DDS} data. For the measurements at \SI{10}{\percent} nominal rate, an average particle rate of \SI{4.84(0.24)e7}{\per\second} with a \SI{5.66(0.14)}{\milli\meter} $\times$ \SI{6.20(0.17)}{\milli\meter} \gls{FWHM} beam spot was measured. The pileup probability in the lower-flux measurements is correspondingly reduced due to the lower average particle rate. The progression of the pileup probability resembles the well-known expression for pileup probability, $1-\exp({-n\cdot \tau})$, derived from Poisson statistics ~\cite{Knoll}, asymptotically approaching \SI{100}{\percent} probability for extended processing times $T$. However, substituting the expectation value $E[\Delta t]$ from inter-arrival time distributions as the average rate $n$ systematically underestimates the pileup observed in all measurements.\\

 The presented measurements of the beam structure can serve as a basis for the design of microdosimetric measurement setups in clinical ion beams, influencing the choice of acceptable shaping times and detector dimensions. In high-resolution amplitude resolved measurements, achieving sufficient charge resolution in pulse height analysis necessitates prolonged shaping times on the order of \SI{1}{\micro\second} ~\cite{Bertuccio_2023_Electronic_Noise}.  To enable single-particle detection at high dose rates and minimize pulse pileup, it is common to reduce the detector cross-section, thereby lowering the average particle rate ~\cite{Knopf_PMB}. However, this method yields only limited suppression of pile-up due to the stochastically distributed particle arrival times. While the average particle rate can be reduced by lowering the cross-sectional area, the probability of pileup remains significant. Even at a low average particle rate, bunching of particles on the $\SI{}{\nano\second}$-scale is observed. These insights should be taken into account for the optimization of readout electronics.\\

In the measurements, a high \gls{SNR} could be achieved for carbon-ion beams and low-energy protons. The detection of high-energy protons was constrained by their low stopping power. Future work should focus on improving the \gls{SNR} of the readout system and investigate the use of thicker detectors to be able to reliably detect high-energy protons. The development of a dedicated multichannel readout to allow for a spatially resolved analysis of the spill structure with an absolute particle count and the use of a time-to-digital converter (TDC) for sub-ns resolution warrants further research. Moreover, it is recommended that the analysis be repeated for different accelerators, including proton cyclotrons, which are widely employed in medical physics research. Nevertheless, the study highlights the potential of SiC as a next-generation detector material, particularly well-suited for high-dose-rate beam monitoring.\\
 
\section{Conclusion}

A SiC-based \gls{HF} readout system was successfully employed for the first time to characterize the microscopic spill structure of clinical ion beams at MedAustron. This emphasizes the feasibility of using SiC detectors for monitoring high flux beams at medical synchrotrons. The \gls{HF} readout is capable of resolving the characteristic spill structure of a medical synchrotron, which is modulated both by ripples from the power converters in the \SI{}{\kilo\hertz} range as well as the synchrotron \gls{RF} frequency in the \SI{}{\mega\hertz} range due to ripple suppression using empty bucket channeling. An analysis over such a wide frequency band is not possible with current beam monitoring systems based on ionization chambers due to limitations in the sampling rate and charge collection time. The simultaneous characterization of both the macroscopic and microscopic spill structure in the iso-centre could potentially serve as input for quality assurance and monitoring beam delivery. The analysis provides valuable insights for research at medical synchrotrons. Due to the stochastic nature of the extraction process, the fluctuations in the particle distribution of the extracted beam obey Poisson statistics ~\cite{PIMMS}. If the spill current is sampled at a high frequency, this nature emerges in measurements leading to the exponential shapes in the inter-arrival time distributions. Due to empty bucket channeling, the inter-arrival times exhibit a modulation leading to non-Poissonian behavior ~\cite{Milosic_2021_Sub_ns}. These considerations are particularly relevant in experimental settings that require reliable single-particle detection with minimal pileup. 

\section*{Acknowledgements}

The financial support of the Austrian Ministry of Education, Science and Research is gratefully acknowledged for providing beam time and research infrastructure at MedAustron. This project has received funding from the Austrian Research Promotion Agency FFG, grant number 918092.

\bibliography{mybibfile}

\end{document}